\pdfoutput=1

\documentclass[11pt]{article}

\usepackage[preprint]{acl}
\usepackage{enumitem}
\usepackage{times}
\usepackage[percent]{overpic}
\usepackage{tabularx}
\usepackage{latexsym}
\usepackage{siunitx}
\usepackage[T1]{fontenc}
\usepackage[utf8]{inputenc}
\usepackage{pmboxdraw} 
\usepackage[T1]{fontenc}
\usepackage{listings}
\usepackage[utf8]{inputenc}
\usepackage{geometry}
\usepackage{enumitem}
\usepackage{color}
\usepackage[utf8]{inputenc}

\usepackage{microtype}

\usepackage{inconsolata}

\usepackage{graphicx}

\usepackage{times}
\usepackage{soul}
\usepackage{url}
\usepackage{amsmath}
\usepackage{amsthm}
\usepackage{booktabs}
\usepackage{algorithm}
\usepackage{algorithmic}
\usepackage[switch]{lineno}
\usepackage{xcolor}
\usepackage{nameref}
\usepackage{microtype}
\usepackage{amsmath}
\usepackage{amsthm,amssymb}
\usepackage{bbm}
\usepackage{graphicx}
\usepackage{booktabs} 
\usepackage{multirow}
\usepackage{acl}
\usepackage{subfigure}
\usepackage{array}
\usepackage{arydshln}
\usepackage{color}
\usepackage{stfloats}
\usepackage{enumitem}

\makeatletter
\newcommand\footnoteref[1]{\protected@xdef\@thefnmark{\ref{#1}}\@footnotemark}
\makeatother

\usepackage{soul}

\definecolor{AddColor}{RGB}{0,130,0}      
\definecolor{ClarifyColor}{RGB}{0,60,160} 
\definecolor{MoveColor}{RGB}{180,90,0}    

\usepackage{xcolor}
\usepackage[linewidth=1pt]{mdframed}

\usepackage{hyperref}
\hypersetup{
    colorlinks=true,
    linkcolor=blue,
    filecolor=magenta,      
    urlcolor=cyan,
    pdftitle={FinGEAR: Financial Mapping-Guided Enhanced Answer Retrieval},
    pdfpagemode=UseOutlines,
    }

\title{FinGEAR: Financial Mapping-Guided Enhanced Answer Retrieval}

\author{
\textbf{Ying Li\textsuperscript{1}} \quad
\textbf{Mengyu Wang\textsuperscript{1}} \quad
\textbf{Miguel de Carvalho\textsuperscript{1,2}} \\
\textbf{Sotirios Sabanis\textsuperscript{1,3,4}} \quad
\textbf{Tiejun Ma\textsuperscript{1,5}}\\
\textsuperscript{1}The University of Edinburgh, United Kingdom\\
\textsuperscript{2}University of Aveiro, Portugal\\
\textsuperscript{3}National Technical University of Athens, Greece\\
\textsuperscript{4}Archimedes/Athena Research Centre, Greece\\
\textsuperscript{5}The Artificial Intelligence Applications Institute, The University of Edinburgh, United Kingdom\\
\texttt{\normalsize \{sunnie.y.li, mengyu.wang, miguel.decarvalho, s.sabanis, tiejun.ma\}@ed.ac.uk}
}

\begin{document}
\maketitle
\begin{abstract}
Financial disclosures such as 10-K filings pose challenging retrieval problems because of their length, regulatory section hierarchy, and domain-specific language, which standard retrieval-augmented generation (RAG) models underuse. We present \textbf{FinGEAR} (Financial Mapping-Guided Enhanced Answer Retrieval), a retrieval framework tailored to financial documents. FinGEAR combines a finance lexicon for Item-level guidance (FLAM), dual hierarchical indices for within-Item search (Summary Tree and Question Tree), and a two-stage cross-encoder reranker. This design aligns retrieval with disclosure structure and terminology, enabling fine-grained, query-aware context selection. Evaluated on full 10-Ks with the FinQA dataset, FinGEAR delivers consistent gains in precision, recall, F1, and relevancy, improving F1 by up to 56.7\% over flat RAG, 12.5\% over graph-based RAGs, and 217.6\% over prior tree-based systems, while also increasing downstream answer accuracy with a fixed reader. By jointly modeling section hierarchy and domain lexicon signals, FinGEAR improves retrieval fidelity and provides a practical foundation for high-stakes financial analysis. 

\end{abstract}

\section{Introduction}
\label{sec:intro}

Financial disclosures such as 10-K filings are key for investment analysis, regulatory monitoring, and risk assessment. They are long (often 100+ pages) and organized by SEC-mandated Items, for example, Item 1 (Business), Item 1A (Risk Factors), Item 7 (Management’s Discussion and Analysis), and Item 8 (Financial Statements). These sections mix narrative text, tables, and footnotes. Many financial NLP tasks, including sentiment analysis, trend detection, entity extraction, risk detection, and question answering, depend on first retrieving the right passages from these filings. Retrieval is difficult because relevant evidence may be spread across multiple Items or years, and domain synonyms (e.g., “sales” vs. “revenue”) and cross-references are common. Therefore, retrieval remains a major bottleneck for current work~\cite{docfinQA,edge2024localglobalgraphrag,selfrag2023,guo2024lightragsimplefastretrievalaugmented}.

Recent efforts such as DocFinQA~\cite{docfinQA} underscore the difficulty of applying question answering to full-length financial filings. Yet these systems typically treat retrieval as a separate external step and rely on fixed-size chunks or off-the-shelf retrievers, without aligning it to the SEC Item hierarchy and the terminology used in financial reports. This reveals a broader limitation: current retrieval methods struggle with hierarchical organization, domain terms, and the need for precise evidence in financial analysis. FinGEAR directly addresses this gap by redefining retrieval as a first-class objective, tailored to the realities of regulatory filings and their analytical use cases.

Recent advances in Large Language Models (LLMs)~\cite{achiam2023gpt, dubey2024llama, wu2023bloomberggpt} and Retrieval-Augmented Generation (RAG)~\cite{lewis2021retrievalaugmentedgenerationknowledgeintensivenlp} have enabled progress in financial document analysis by grounding outputs in retrieved evidence. Our analysis identifies three core limitations in current retrieval pipelines that constrain downstream performance across financial NLP tasks:
(1) \textit{Lack of structure awareness}: fixed-size segmentation discards the logical hierarchy of disclosures, leading to misaligned context retrieval;
(2) \textit{Lack of financial specificity}: generic retrievers fail to distinguish nuanced but crucial concepts (e.g., “net income” vs. “operating income”);
(3) \textit{Dense-only retrieval is hard to control and explain}: pure vector similarity offers limited interpretability in evidence-heavy settings.

To address these issues, we present \textbf{FinGEAR}, \textbf{Fin}ancial Mapping-\textbf{G}uided \textbf{E}nhanced \textbf{A}nswer \textbf{R}etrieval, a retrieval-centric framework designed for long, professionally authored, semi-structured disclosures. FinGEAR treats retrieval as the core problem, aiming to surface content that is structurally coherent, financially grounded, and useful across tasks.

FinGEAR introduces three key contributions:
(1) \textit{Document–Query hierarchical alignment}, which captures the structural layout of financial documents via a Summary Tree and enables query-sensitive retrieval through a structurally mirrored Question Tree;
(2) \textit{Financial Lexicon-Aware Mapping (FLAM)}, which steers retrieval using domain-specific term clusters and lexicon-weighted scoring;
(3) \textit{Hybrid dense–sparse retrieval}, which integrates sparse keyword anchoring with dense embedding similarity to balance interpretability and relevance.

Evaluated on full 10-K filings, FinGEAR achieves up to 138\% higher retrieval F1 than flat RAG, up to 28\% over graph-based baselines (e.g., LightRAG), and up to 263\% over prior tree-based systems. Ablation studies confirm that these gains derive from the combined design of its structural and domain-aware modules. While FinGEAR does not directly optimize for reasoning tasks, downstream experiments confirm that enhanced retrieval leads to better answer accuracy, reinforcing retrieval quality as the foundation of financial document understanding.

To our knowledge, FinGEAR is the first retrieval-first system tailored to financial disclosures. It offers a principled and modular foundation for structured, explainable, and task-flexible financial NLP.
\section{Related Work}
\label{sec:related_work}

\subsection{Retrieval-Augmented Generation (RAG)}

Retrieval-Augmented Generation (RAG)~\cite{lewis2021retrievalaugmentedgenerationknowledgeintensivenlp} augments language models by fetching relevant context from external corpora, reducing the need for full-model fine-tuning~\cite{guu2020retrievalaugmented, ram2023}. Advanced variants such as Self-RAG~\cite{selfrag2023} and Adaptive RAG~\cite{adaptive2023} improve coordination between retrievers and generators but still use fixed-size chunks. This makes it hard to preserve document structure and can introduce drift in long documents, as seen in long-form QA benchmarks like ELI5~\cite{fan-etal-2019-eli5}. Recent work addresses context-length limits with longer-context models (e.g., Transformer-XL~\cite{dai2019transformerxlattentivelanguagemodels}), retrieval-aware chunking~\cite{zhong2025mixofgranularityoptimizechunkinggranularity}, and studies of position bias~\cite{liu2023lostmiddlelanguagemodels}. These efforts mainly target input length and do not solve structured retrieval in financial domains.

\subsection{Hierarchical and Graph-Based Retrieval}

Hierarchical methods such as RAPTOR~\cite{raptor} and HiQA~\cite{chen2024hiqa} represent documents as trees and retrieve recursively from higher-level summaries. Graph-based systems, including GraphRAG~\cite{edge2024localglobalgraphrag} and LightRAG~\cite{guo2024lightragsimplefastretrievalaugmented}, model relations between entities and sections to support multi-hop reasoning. In particular, GraphRAG builds local–global graphs over LLM-extracted entities and community summaries and retrieves via community-level traversal, while LightRAG performs dual-level query decomposition with lightweight neighborhood expansion over section-aligned segments. Longtriever~\cite{yang-etal-2023-longtriever} targets long-context retrieval by combining local and global semantics at the block level. These graph approaches differ from retrieval that uses section hierarchies (e.g., SEC Items), so we include GraphRAG as a representative graph baseline for community-level traversal. Many graph/tree systems also depend on LLM-generated summaries, which may hallucinate content~\cite{maynez2020faithfulnessfactualityabstractivesummarization, li2023extracting}.

\subsection{Financial NLP and Domain-Specific Retrieval}

Financial NLP supports applications such as sentiment analysis~\cite{zhang2023financialsentiment, wang2024mana}, event prediction~\cite{li2024investorbench, wang2024modeling}, and hybrid QA~\cite{chen2021finqa, zhu2021tatqaquestionansweringbenchmark}, often using domain-adapted models like FinBERT-QA~\cite{finbertqa} and FinGPT~\cite{yang2023fingpt}. These models improve semantic understanding but usually assume that relevant context is provided and therefore lack retrieval. DocFinQA~\cite{docfinQA} evaluates QA over full filings but relies on an oracle retriever, leaving retrieval design unaddressed. As a result, prior work does not fully model retrieval architectures that reflect hierarchical layouts, domain terminology, and section-specific semantics. FinGEAR addresses this gap by treating structure-aware retrieval as a core objective in financial document understanding.

\begin{figure*}[t]
\centering
\includegraphics[width=\linewidth]{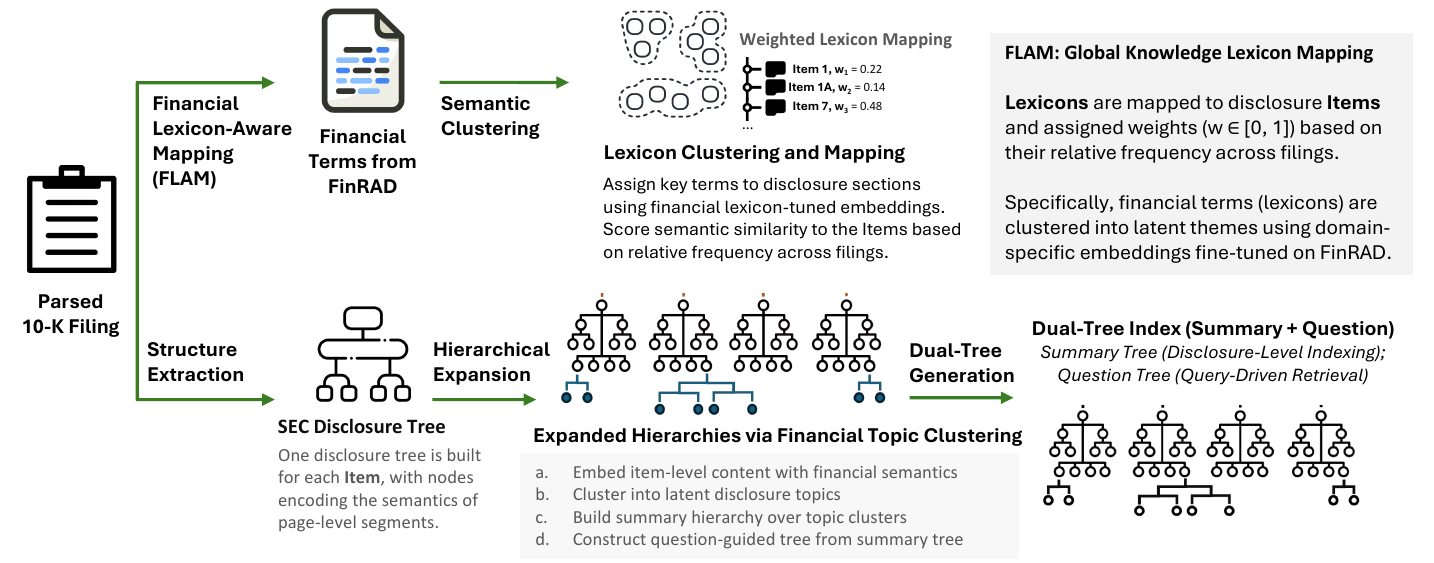}
\vspace{-0.8em}
\caption{Pre-retrieval. From parsed 10-K filings, FinGEAR performs structure extraction and lexicon mapping (FLAM). FLAM clusters domain terms and assigns Item weights; topic clustering builds a Summary Tree and a mirrored Question Tree for each Item.}
\label{fig:pre_retrieval}
\vspace{-0.5em}
\end{figure*}

\subsection{Guided and Interpretable Retrieval for Financial Documents}

In high-stakes settings like finance, retrieval should be both relevant and interpretable because results support regulatory or analytical decisions~\cite{yu2024defense}. Flat, dense-only pipelines can obscure why passages were selected. Structured methods (hierarchical and graph-based) reviewed above improve traceability by encoding document structure and relations, and they typically outperform flat chunking in both quality and transparency. However, most remain domain-agnostic. For 10-Ks, where a standardized section layout and stable terminology are available, integrating domain signals (e.g., a finance lexicon and disclosure Item hierarchy) can further align retrieval with analyst intent. FinGEAR follows this principle by combining lexicon-guided global navigation with Item-aligned hierarchical indexing, providing interpretable, section-aware evidence selection tailored to financial disclosures.

\begin{figure*}[t]
\centering
\includegraphics[width=\linewidth]{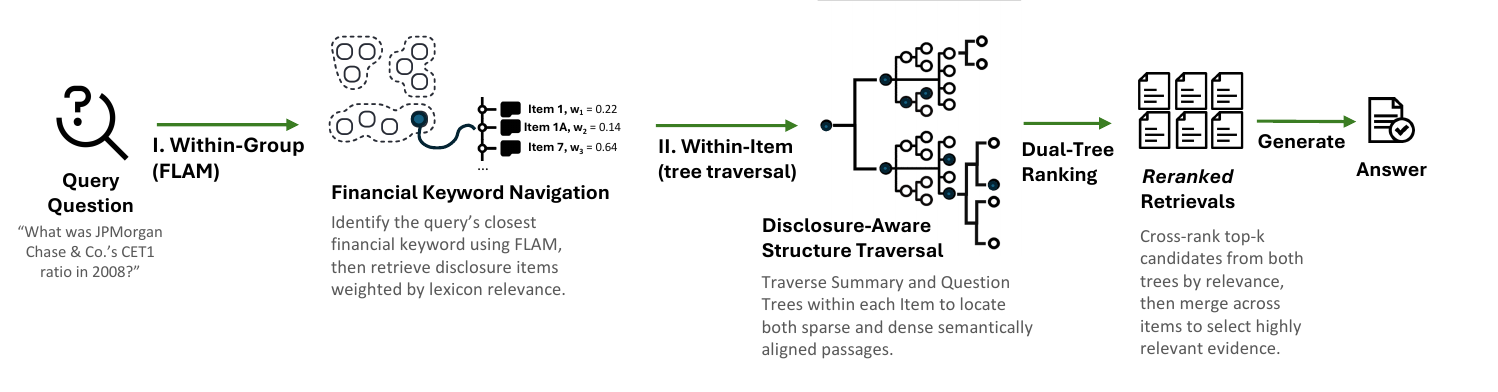}
\vspace{-1.2em}
\caption{In-retrieval. FLAM allocates the budget across Items (Within-Group). Within each Item, the Summary and Question Trees are traversed (Within-Item). Candidates are jointly reranked and merged across Items. Example query: CET1 ratio in 2008.}
\label{fig:in_retrieval}
\vspace{-0.8em}
\end{figure*}
\section{Methodology}
\label{sec:methodology}

FinGEAR is a modular retrieval framework designed to align with the structure and terminology of long financial filings. It is motivated by the observation that 10-K reports embed rich domain-specific signals, such as SEC Item headings, disclosure hierarchy, and financial terms. These signals can be extracted and used for improved retrieval. Rather than relying on flat chunking or dense-only similarity, FinGEAR builds hierarchical representations and uses a financial lexicon to guide search with hybrid matching, enabling retrieval with structural fidelity and domain specificity.

\subsection{Pre-Retrieval Pipeline}

Before retrieval, FinGEAR builds indexing structures for context-aware navigation: (1) \textit{structure extraction} to model disclosure hierarchy, and (2) \textit{Financial Lexicon-Aware Mapping (FLAM)} to steer search toward financially salient content.
Figure~\ref{fig:pre_retrieval} summarizes this pre-retrieval stage, showing dual-tree construction (Summary/Question Trees) alongside FLAM’s lexicon-to-Item weighting that will drive per-Item budgets used later (Section~\ref{sec:in_retrieval}).

\begin{figure}[t]
\centering
\includegraphics[width=0.9\linewidth]{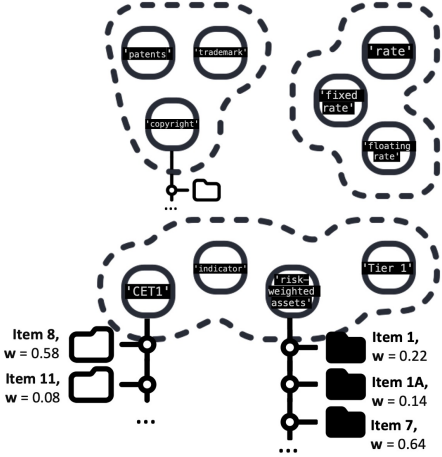}
\vspace{-0.6em}
\caption{Global navigation with FLAM. Lexicon clusters map query terms to disclosure Items and assign weights $w\!\in\![0,1]$ based on relative frequency across filings; these weights determine per-Item budgets $k_i^{\ast}$ used for traversal.}
\label{fig:flam_weights}
\vspace{-0.8em}
\end{figure}

\subsubsection{Structure Extraction}
\label{sec:structure_extraction}

FinGEAR extracts fine-grained structure within each 10-K Item. While SEC-mandated itemization provides a high-level layout~\cite{alberg1970sec}, each Item can span dozens of pages and include heterogeneous content. To support retrieval at multiple granularities, FinGEAR builds semantic trees within each Item using topic-based clustering. This hierarchical organization is depicted in Figure~\ref{fig:pre_retrieval} as the Summary/Question Tree index built during pre-retrieval.

\paragraph{Chunking and Encoding}
Each Item is segmented into approximately 2{,}000-token chunks with a 100-token overlap (see Appendix~\ref{appendix:token_statistics}, Table~\ref{tab:token_statistics}). We encode each chunk with sentence embeddings fine-tuned on financial data, yielding a domain-aligned representation space.

\begin{figure}[t]
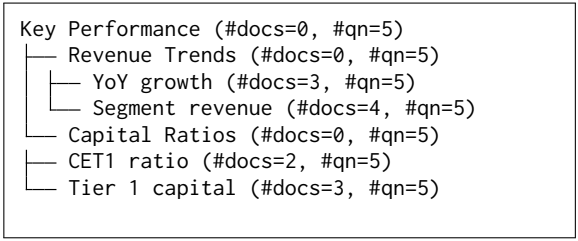

\centering
\setlength{\fboxsep}{6pt}
\fbox{%
\begin{minipage}{0.92\linewidth}
\ttfamily\small
Key Performance (\#docs=0, \#qn=5)\\
├── Revenue Trends (\#docs=0, \#qn=5)\\
│   ├── YoY growth (\#docs=3, \#qn=5)\\
│   └── Segment revenue (\#docs=4, \#qn=5)\\
└── Capital Ratios (\#docs=0, \#qn=5)\\
    ├── CET1 ratio (\#docs=2, \#qn=5)\\
    └── Tier 1 capital (\#docs=3, \#qn=5)\\
\end{minipage}}
\caption{Example tree printout. Internal nodes are summaries; leaves are retrievable chunks. “\#qn” is the number of sub-questions stored at the corresponding Question Tree node.}
\label{fig:tree_printout}
\vspace{-0.6em}
\end{figure}

\paragraph{Hierarchical Expansion via Topic Clustering}
We then build the hierarchy bottom–up: UMAP is first applied for dimensionality reduction, followed by Gaussian Mixture Models (GMM) for soft clustering. Leaves correspond to the original chunks; each internal node represents a soft cluster and is summarized to guide top-down traversal. The result is a content hierarchy that captures themes from coarse to fine granularity (hyperparameters in Section~\ref{sec:experiments_setup}; schematic in Figure~\ref{fig:pre_retrieval}). An example of the resulting hierarchy is shown in Figure~\ref{fig:tree_printout}.

\paragraph{Summary and Question Trees (Outputs)}
Structure extraction yields two structurally identical hierarchical indices. The \textit{Summary Tree} stores node summaries at internal nodes and the original text chunks at leaves. The \textit{Question Tree} mirrors the same topology but stores LLM-generated sub-questions (embedded in the same query space) at internal nodes; its leaves point to the same chunk IDs as the Summary Tree. We derive the Question Tree by generating sub-questions for each Summary-Tree node and embedding them with a FinQA-aligned encoder. Sharing topology while differing in node content enables hybrid sparse–dense traversal during retrieval. Fine-tuning and further construction details appear in Appendix~\ref{appendix:embedding_finetuning}.

\subsubsection{Financial Lexicon-Aware Mapping (FLAM)}

FLAM provides domain-aware guidance by weighting Items before traversal. Candidate terms are extracted using a rule-based method~\cite{singh2017replaceretrievekeywordsdocuments} from a curated subset of the FinRAD lexicon~\cite{ghosh2021finrad}, and then clustered with sentence embeddings fine-tuned on financial language (Appendix~\ref{appendix:embedding_finetuning}). Each relevant term is assigned a weight using \textit{Relative Frequency}:
\[
\text{weight}(k_i) = \frac{\text{count}(k_i)}{\sum_j \text{count}(k_j)},
\]
chosen for its interpretability and robustness across heterogeneous filings. (Ablations appear in Section~\ref{sec:results} and Section~\ref{sec:ablation_study}.)

\vspace{0.35\baselineskip}
\noindent\textbf{FLAM vs.\ structure extraction.}
FLAM operates at the corpus level: it clusters \emph{lexical terms} (FinRAD-tuned embeddings) and converts them into per-Item \emph{weights} that allocate the global retrieval budget.
Structure extraction operates within each 10-K Item: it clusters \emph{text chunks} (FinQA-tuned embeddings) to build \emph{local trees} used for traversal.
In short, FLAM decides \emph{where} to look across Items, while structure extraction decides \emph{how} to search within an Item.

\subsection{In-Retrieval Pipeline}
\label{sec:in_retrieval}

At query time, FLAM first identifies Items that are likely to contain relevant content. Within these Items, dual-tree traversal retrieves candidate passages: the \textit{Summary Tree} provides sparse, high-level routing; the \textit{Question Tree} provides dense, query-specific refinement. Figure~\ref{fig:in_retrieval} illustrates this process for an example FinQA query: \emph{“What was JPMorgan Chase \& Co.’s CET1 ratio in 2008?”}.

\subsubsection{Global Navigation}

Global Navigation selects which SEC Items to search before any tree traversal. Using FLAM, we (i) expand the query into clusters of related financial terms, (ii) measure how strongly each Item is associated with those clusters to obtain normalized weights \(w_i\), and (iii) convert the weights into per-Item retrieval budgets \(k_i^{\ast}\) such that \(\sum_i k_i^{\ast}=k\). Items with \(k_i^{\ast}=0\) are skipped; the rest proceed to within-Item traversal.

\noindent\textbf{Procedure.}
\begin{enumerate}[leftmargin=*, itemsep=2pt, topsep=2pt]
    \item \textit{Term detection and expansion.} Extract salient financial terms from the query (e.g., \textit{CET1}, \textit{capital ratio}) and expand them using FLAM’s lexicon clusters (FinRAD-tuned embeddings) to capture close variants.
    \item \textit{Map terms to Items.} For each expanded term cluster, count its occurrences in each SEC Item across the recovered filings and aggregate these counts per Item. Convert counts to normalized Item weights
    \[
      w_i \;=\; \frac{\mathrm{freq}_i}{\sum_j \mathrm{freq}_j}\,,\qquad \sum_i w_i = 1,\; w_i \in [0,1].
    \]
    \item \textit{Allocate the budget.} Given a total retrieval budget \(k\), assign an Item-level budget
    \[
      k_i^{\ast} \;=\; \mathrm{round}\!\big(k \cdot w_i\big)
    \]
    (with a final adjustment so that \(\sum_i k_i^{\ast}=k\); ties are broken by larger \(w_i\)).
    \item \textit{Handoff to traversal.} For each Item with \(k_i^{\ast}>0\), pass the Item and its budget to the within-Item search over the Summary and Question Trees.
\end{enumerate}

\noindent Item weights \(w_i\) (e.g., \(w_{1}\), \(w_{1\mathrm{A}}\), \(w_{7}\)) are illustrated in Figure~\ref{fig:flam_weights} to match the notation used in Section~\ref{sec:in_retrieval}. For example, with \(k{=}10\): Item~7 receives \(k^{\ast}{=}6\), Item~1 \(k^{\ast}{=}2\), and Item~1A \(k^{\ast}{=}2\).

\subsubsection{Within-Item Search}

Within each selected Item, we traverse two trees with identical topology, the \emph{Summary Tree} and the \emph{Question Tree}. The structure (parent–child layout and leaf set) is the same; what differs is the node content used for scoring during traversal: Summary-Tree nodes carry summaries, while Question-Tree nodes carry LLM-generated sub-questions embedded in the query space. Leaf nodes in both trees reference the same chunk IDs, so leaf hits are merged and deduplicated into a single candidate pool.

\begin{itemize}
    \item \textit{Summary-based retrieval (sparse):} nodes are matched to the query using a bag-of-words scorer over node \emph{summaries}. This favors sections dense in relevant financial terms and headings.
    \item \textit{Question-based retrieval (dense):} nodes are matched using cosine similarity between the query embedding and LLM-generated \emph{sub-questions} stored at each node (financial QA–tuned embeddings).
\end{itemize}

\noindent\textit{Scoring and stopping.} The two trees are traversed independently. In the Summary Tree, nodes are scored with BM25 over node summaries (sparse signal). In the Question Tree, nodes are scored by cosine similarity between the query and each node’s sub-questions (dense signal). At each internal node we expand only the top \(b\) children by that node’s score (default \(b{=}3\) in all experiments). This per-node child limit is separate from the per-Item budget \(k_i^{\ast}\). We continue descending until leaves. Because the trees are built with depth at most 2 (Section~\ref{sec:experiments_setup}), traversal reaches leaves in at most two steps. All leaves visited across both trees form the Item’s candidate pool, from which we select up to \(k_i^{\ast}\) evidence chunks.

\paragraph{Reranking.}
\textbf{Stage 1 (cross-tree):} merge and rerank candidates from the Summary and Question traversals with a cross-encoder (\emph{BAAI/bge-reranker-large}, XLM-R-Large, 560M).\\
\textbf{Stage 2 (cross-Item):} rerank the top spans across different Items to prioritize globally informative, coherent answers.
This two-stage reranking sharpens precision while preserving coverage for downstream QA.

Finally, we select up to \(k_i^{\ast}\) highest-scoring unique chunks for that Item and pass them to reranking.
\section{Dataset and Evaluation}
\label{sec:dataset_metrics}

\subsection{Datasets}

To evaluate our retrieval capabilities, we use the FinQA dataset~\cite{chen2021finqa}, a benchmark designed for financial question answering. FinQA contains 8{,}281 question–answer pairs derived from 10-K filings, requiring numerical reasoning and domain expertise. The dataset is split into training (6{,}251), validation (883), and testing (1{,}147) sets.

While the FinQA dataset provides ground-truth question–answer pairs, it only includes \textit{pre-extracted} context passages. To enable full-document retrieval evaluation, we recover the original 10-K filings corresponding to each FinQA instance. Using SEC EDGAR records\footnote{\url{https://www.sec.gov/search-filings}}, we match company names to ticker symbols and Central Index Keys (CIKs), resulting in a corpus of 720 full-length 10-K filings (we refer to these as our \textit{recovered 10-K filings}).

These documents span a diverse set of industries—including technology, healthcare, financial services, consumer goods, and energy—and cover companies listed in the S\&P 500 index between 1999 and 2019. Filings are converted from PDF to structured Markdown format, preserving original SEC itemization. This setup ensures FinGEAR retrieves from complete, uncurated documents, reflecting real-world retrieval challenges.

\noindent\textit{Dataset breadth.} FinQA dataset includes both numerical and categorical questions and covers single-step and multi-hop reasoning. We use this typology in our evaluations (see Section~\ref{sec:results} and Table~\ref{tab:dataset_ablation_results}), underscoring that FinQA dataset is a multifaceted dataset rather than a narrowly scoped benchmark.

\subsection{Evaluation Framework}
\label{subsec:evaluation_framework}

We evaluate FinGEAR’s retrieval performance using the RAGAS framework~\cite{shahul2023ragas}, which provides \textit{component-level RAG evaluation metrics}. Specifically, we report:

\noindent\textbf{Precision.} The proportion of retrieved passages that are relevant to the query. \\
\textbf{Recall.} The proportion of all relevant passages that are successfully retrieved. \\
\textbf{F1 Score.} The harmonic mean of precision and recall. This serves as our \textit{primary} retrieval metric, as it best isolates retrieval quality without conflating it with downstream generation performance. \\
\textbf{Relevancy.} A semantic alignment score based on LLM evaluations, measuring how well retrieved passages support the query’s intent.

Retrieval is evaluated at multiple depths, including Top-5, Top-10, and Top-15, to capture trade-offs between precision and coverage. All metrics are reported per depth and averaged to provide a comprehensive assessment.

In addition to retrieval metrics, we report \textit{final answer accuracy}, defined as the correctness of a fixed reader model’s response given the retrieved context (reader details in Section~\ref{sec:experiments_setup}). While not a direct measure of retrieval quality, it reflects the downstream utility of the system for real-world financial question answering and analytical tasks.

To improve transparency, we also report token-level statistics in Appendix~\ref{appendix:token_statistics}, including average tokens per passage, total input length by depth, average tree depth, and maximum context constraints.
\section{Experiments}
\label{sec:experiments}

\subsection{Experimental Setup}
\label{sec:experiments_setup}

\begin{table*}[htbp!]
    \centering
    \caption{Retrieval performance comparison across baseline models. Results are reported at retrieval depths \(k = 5, 10, 15\), where \(k\) denotes the number of retrieved passages. The best score for each metric is shown in \textbf{bold}, and the second-best is \underline{underlined}.}
    \label{tab:ragas_baselines}
    \resizebox{\textwidth}{!}{
    \begin{tabular}{l|ccc|ccc|ccc|ccc}
        \toprule
        \multirow{1}{*}{\textbf{Model}} & \multicolumn{3}{c}{\textbf{Precision}} & \multicolumn{3}{c}{\textbf{Recall}} & \multicolumn{3}{c}{\textbf{F1 Score}} & \multicolumn{3}{c}{\textbf{Relevancy}}  \\
        & \(k=5\) & \(k=10\) & \(k=15\) & \(k=5\) & \(k=10\) & \(k=15\) & \(k=5\) & \(k=10\) & \(k=15\) & \(k=5\) & \(k=10\) & \(k=15\)\\
        \midrule
        General RAG  & 0.37 & 0.37 & 0.30 & 0.24 & 0.26 & 0.28 & 0.29 & 0.30 & 0.29 & \underline{0.40} & \underline{0.43} & \underline{0.47} \\
        Self-RAG     & 0.74 & 0.60 & 0.55 & 0.27 & 0.28 & 0.31 & 0.39 & 0.38 & 0.40 & 0.30 & 0.31 & 0.33 \\
        LightRAG     & \textbf{0.88} & \underline{0.85} & \underline{0.85} & 0.39 & 0.42 & 0.47 & 0.54 & 0.56 & 0.60 & 0.38 & 0.37 & 0.39\\
        GraphRAG & \textbf{0.88} & \textbf{0.89} & \textbf{0.87} & \underline{0.56} & \underline{0.55} & \underline{0.55} & \underline{0.67} & \underline{0.66} & \underline{0.66} & 0.17 & 0.16 & 0.17 \\
        RAPTOR       & 0.69 & 0.65 & 0.62 & 0.11 & 0.14 & 0.22 & 0.19 & 0.23 & 0.32 & 0.38 & 0.41 & 0.45\\
        \textbf{FinGEAR} & \underline{0.79} & 0.76 & 0.72 & \textbf{0.61} & \textbf{0.62} & \textbf{0.65} & \textbf{0.69} & \textbf{0.68} & \textbf{0.68} & \textbf{0.50} & \textbf{0.64} & \textbf{0.62}\\
        \bottomrule
    \end{tabular}
    }
\end{table*}

\paragraph{Baselines.}
We benchmark against \textit{General RAG}~\cite{lewis2021retrievalaugmentedgenerationknowledgeintensivenlp}, \textit{Self-RAG}~\cite{selfrag2023}, \textit{LightRAG}~\cite{guo2024lightragsimplefastretrievalaugmented}, \textit{GraphRAG}~\cite{edge2024localglobalgraphrag}, and \textit{RAPTOR}~\cite{raptor}.
LightRAG is run in its “mix” mode (keyword+dense with local neighborhood expansion).
GraphRAG follows the official local–global community summary traversal (community construction and inference-time community propagation).
Unless noted, baselines use default settings with \texttt{text-embedding-ada-002} for dense similarity.
(Full baseline configurations are summarized in Appendix~\ref{appendix:baseline_configurations}.)

\paragraph{Reader model.}
All results use \texttt{GPT-4o-mini} as a fixed reader with default settings; only the retriever varies across systems. It was chosen for being a strong, cost- and latency-efficient reader at the time of this study. Crucially, the reader choice is independent of the retrieval pipeline and algorithmic design as our conclusions target retrieval quality.

\paragraph{Model and index settings (promoted from appendices).}
\textbf{BM25:} Lucene; \(k_1{=}1.5\), \(b{=}0.75\); tokenizer: PyStemmer (eng). \\
\textbf{Dense embeddings:} \texttt{BAAI/bge-base-en-v1.5} (\(d{=}768\)); fine-tuned FinRAD \& FinQA variants with MultipleNegatives + Matryoshka; learning rate \(2{\times}10^{-5}\); batch size 32. Empirical gains from the embedding fine-tuning are summarized in Table~\ref{tab:embedding_finetuning}.\\
\textbf{Cross-encoder reranker:} \texttt{BAAI/bge-reranker- large} (default inference settings). \\
\textbf{UMAP+GMM:} UMAP dim{=}10 (cosine); GMM max components{=}50; threshold{=}0.1; max tree depth{=}2 (fan-out rationale from Item sizes). \\
\textbf{Chunking:} \(\sim\)2{,}000 tokens with 100-token overlap.

\paragraph{Retrieval Pipeline.}
We use \textbf{BM25s}~\cite{bm25s} for sparse matching and \textbf{FinLang} domain-specific sentence embeddings~\cite{finlang2025finance} for dense similarity. The embeddings are fine-tuned on FinQA (question–passage) and FinRAD (lexicon–sentence) objectives (Appendix~\ref{appendix:embedding_finetuning}). Candidate financial terms are extracted with spaCy’s \texttt{PhraseMatcher} to drive FLAM’s term clustering and Item weighting. Unless noted otherwise, all experiments use this BM25s+FinLang configuration; hierarchical indexing and reranking are detailed in Section~\ref{sec:methodology}.

\paragraph{Retrieval Settings.}
Retrieval performance is reported at depths \(k = 5, 10, 15\), reflecting trade-offs between retrieval quantity and contextual precision. All models are evaluated using identical traversal logic and input constraints to ensure comparability. Additional configuration details, including chunk size, overlap settings, and node budget, are reported in Appendix~\ref{appendix:token_statistics}, Table~\ref{tab:token_statistics}.

\paragraph{Ablation Settings.} 
To assess the contribution of FinGEAR’s core components, we conduct a series of ablation studies. We begin by disabling individual modules, the \textit{Summary Tree}, \textit{Question Tree}, and \textit{FLAM}, to measure their impact on retrieval F1 and Relevancy. We then perform multi-component ablations by jointly removing pairs of modules to analyze interaction effects between structural and lexical guidance (Appendix~\ref{appendix:multi_ablation}).
In addition, we evaluate alternative lexicon weighting strategies within the FLAM module, including \textit{Relative Frequency}, \textit{Exponential Scaling}, and \textit{Softmax Weighting}. These modules affect Item prioritization during retrieval. Ablation results are presented in Section~\ref{sec:results}.

\subsection{Results}
\label{sec:results}

\begin{table*}[htbp!]
    \centering
    \caption{Ablation study on the impact of removing individual components from FinGEAR. 
    We evaluate the effect of disabling the Summary Tree, Question Tree, FLAM modules and Reranker. 
    Results are reported at retrieval depths \(k = 5, 10, 15\) and averaged across metrics. 
    \textbf{Bold} indicates the best-performing configuration.}
    \label{tab:ablation_single_component}
    \resizebox{\textwidth}{!}{
    \begin{tabular}{l|ccc|ccc|ccc|ccc}
        \toprule
        \multirow{1}{*}{\textbf{Ablation Setting}} & \multicolumn{3}{c}{\textbf{Precision}} & \multicolumn{3}{c}{\textbf{Recall}} & \multicolumn{3}{c}{\textbf{F1 Score}} & \multicolumn{3}{c}{\textbf{Relevancy}}\\
        & \(k=5\) & \(k=10\) & \(k=15\) & \(k=5\) & \(k=10\) & \(k=15\) & \(k=5\) & \(k=10\) & \(k=15\) & \(k=5\) & \(k=10\) & \(k=15\) \\
        \midrule
        \textbf{Full FinGEAR}     & \textbf{0.79} & \textbf{0.76} & \textbf{0.72} & \textbf{0.61} & \textbf{0.62} & \textbf{0.65} & \textbf{0.69} & \textbf{0.68} & \textbf{0.68} & \textbf{0.50} & \textbf{0.64} & \textbf{0.62} \\
        No Summary Tree           & 0.41 & 0.61 & 0.58 & 0.33 & 0.36 & 0.40 & 0.37 & 0.46 & 0.47 & 0.29 & 0.57 & 0.63 \\
        No Question Tree          & 0.43 & 0.62 & 0.57 & 0.35 & 0.37 & 0.40 & 0.39 & 0.47 & 0.47 & 0.29 & 0.56 & 0.63 \\
        No FLAM Module            & 0.42 & 0.61 & 0.58 & 0.34 & 0.35 & 0.42 & 0.38 & 0.44 & 0.49 & 0.29 & 0.58 & 0.63 \\
        No Reranker            & 0.51 & 0.44 & 0.36 & 0.37 & 0.30 & 0.31 & 0.43 & 0.36 & 0.34 & 0.45 & 0.55 & 0.58 \\
        \bottomrule
    \end{tabular}
    }
\end{table*}

We evaluate FinGEAR’s performance on four core retrieval metrics—\textit{precision}, \textit{recall}, \textit{F1 score}, and \textit{relevancy}—alongside ablation studies and lexicon-weighting variants. While FinGEAR is retrieval-first, we also report \textit{answer accuracy} to assess how improved retrieval supports downstream tasks.

\vspace{-0.5em}
\begin{table}[h]
\centering
\resizebox{\columnwidth}{!}{
\begin{tabular}{lccc}
\toprule
\textbf{System} & \(k=5\) & \(k=10\) & \(k=15\) \\
\midrule
General RAG & 29.8\% & 30.3\% & 30.5\% \\
Self-RAG & 28.7\% & 29.8\% & 27.4\% \\
LightRAG & 35.7\% & 58.8\% & 36.5\% \\
GraphRAG & 28.4\% & 29.1\% & 29.4\% \\ 
RAPTOR & 34.0\% & 20.9\% & 37.3\% \\
\textbf{FinGEAR (Ours)} & \textbf{49.1\%} & \textbf{49.7\%} & \textbf{50.0\%} \\
\bottomrule
\end{tabular}
} 
\caption{Final answer accuracy with GPT-4o-mini as reader. Accuracy is reported for each retrieval depth.}
\label{tab:answer_accuracy}
\end{table}
\vspace{1em}

\subsubsection{Retrieval Performance Across Baselines}

Table~\ref{tab:ragas_baselines} compares FinGEAR with five baselines: General RAG, Self-RAG, LightRAG, GraphRAG, and RAPTOR. Across all retrieval depths and evaluation metrics, including precision, recall, F1 score, and relevancy, FinGEAR consistently outperforms the baselines, demonstrating its effectiveness for structure-aware retrieval over disclosures. 

\textit{Recall.} FinGEAR achieves the highest recall at every depth, reaching 0.65 at \(k{=}15\), ahead of LightRAG (0.47), Self-RAG (0.31), and General RAG (0.28). RAPTOR lags behind, highlighting limited adaptability to the standardized section hierarchy and terminology of 10-Ks. Strong recall indicates that FinGEAR reliably recovers the reasoning spans dispersed across long documents.

\textit{F1.} FinGEAR also leads on F1 at all depths, reflecting a balanced trade-off between precision and coverage. The combination of FLAM-guided Item selection and dual-tree traversal yields candidate sets that are both relevant and diverse—important in high-stakes financial analysis.

\textit{Relevancy.} FinGEAR delivers the most semantically aligned results at all depths, peaking at 0.64. Retrieved passages are not only topically related but also faithful to the specific intent of the query. By contrast, General RAG and RAPTOR return more diffuse or off-target content.

\textit{LightRAG.} LightRAG attains the highest precision in isolation (e.g., 0.88 at \(k{=}5\)), but this comes with reduced recall and F1, reflecting a narrower candidate set after entity/relation hops. FinGEAR maintains competitive precision while substantially improving recall and F1, yielding a more balanced profile for long, heterogeneous filings.\footnote{In LightRAG, each retrieval combines entities, relations, and vector text, which can inflate precision but requires approximately fifteen times the runtime of FinGEAR per document.}

\textit{GraphRAG.} GraphRAG shows high precision and strong recall (0.56/0.55/0.55), yielding second-best F1 (0.67/0.66/0.66). However, its relevancy is lowest (0.17/0.16/0.17), and its final answer accuracy is also low (Table~\ref{tab:answer_accuracy}: 28.4–29.4\%). In our setting, FinGEAR achieves higher F1 and much higher relevancy at all depths, aligning with its stronger downstream accuracy.

\begin{table*}[htbp!]
    \centering
    \caption{
    Ablation study of lexicon weighting strategies in FLAM. 
    We compare three methods for computing weights from financial terminology: Relative Frequency, Logarithmic Weighting, and Softmax Weighting. 
    Scores are reported at retrieval depths \(k = 5, 10, 15\) and averaged across all metrics. 
    \textbf{Bold} indicates the best-performing configuration.
    }
    \label{tab:flam_weighting}
    \resizebox{\textwidth}{!}{
    \begin{tabular}{l|ccc|ccc|ccc|ccc}
        \toprule
        \multirow{1}{*}{\textbf{Weighting Strategy}} & \multicolumn{3}{c}{\textbf{Precision}} & \multicolumn{3}{c}{\textbf{Recall}} & \multicolumn{3}{c}{\textbf{F1 Score}} & \multicolumn{3}{c}{\textbf{Relevancy}}\\
        & \(k=5\) & \(k=10\) & \(k=15\) & \(k=5\) & \(k=10\) & \(k=15\) & \(k=5\) & \(k=10\) & \(k=15\) & \(k=5\) & \(k=10\) & \(k=15\) \\
        \midrule
        \textbf{Relative Frequency} & \textbf{0.79} & \textbf{0.76} & \textbf{0.72} & \textbf{0.61} & \textbf{0.62} & \textbf{0.65} & \textbf{0.69} & \textbf{0.68} & \textbf{0.68} & \textbf{0.50} & \textbf{0.64} & 0.62  \\
        Logarithmic Weighting         & 0.70 & 0.66 & 0.63 & 0.47 & 0.45 & 0.45 & 0.56 & 0.53 & 0.52 & \textbf{0.50} & 0.60 & 0.60 \\
        Softmax Weighting           & 0.70 & 0.68 & 0.66 & 0.47 & 0.44 & 0.44 & 0.56 & 0.53 & 0.53 & 0.48 & 0.59 & \textbf{0.63}\\
        \bottomrule
    \end{tabular}
    }
\end{table*}

\subsubsection{Evaluating QA as a Downstream Task: Answer Accuracy}
To assess downstream utility, we measure final answer accuracy with the fixed reader introduced above. FinGEAR achieves the strongest accuracy at \(k{=}5\) and \(k{=}15\) (49.1\% and 50.0\%), and remains competitive at \(k{=}10\) (49.7\%), where LightRAG shows a one-off spike (58.8\%) but degrades at the other depths (35.7\% at \(k{=}5\), 36.5\% at \(k{=}15\)). General RAG and GraphRAG hover near 30\% across depths, and RAPTOR is unstable, dropping from 34.0\% at \(k{=}5\) to 20.9\% at \(k{=}10\), then rebounding to 37.3\% at \(k{=}15\). Overall, FinGEAR’s accuracy is consistently high and increases with \(k\), mirroring its recall/F1 trends and indicating that it surfaces numerically faithful, context-grounded evidence for fact-sensitive financial QA.

\subsubsection{Analysis of Retrieval Depths}

FinGEAR maintains strong retrieval quality across all evaluated depths. At \(k=5\), it achieves a high precision of 0.79 and F1 score of 0.69, indicating strong early-stage retrieval performance. As retrieval depth increases, recall steadily improves, reaching 0.65 at \(k=15\), while precision and F1 remain consistently high. This stability reflects FinGEAR’s ability to scale retrieval scope without sacrificing semantic accuracy or contextual fit.

Relevancy scores follow a similar pattern—peaking at 0.64 at \(k=10\) and holding at 0.62 at \(k=15\), showing that even as the system retrieves more passages, it continues to surface content that is contextually faithful to the query. These trends suggest that FinGEAR balances depth and alignment effectively, making it suitable for long-context applications where both coverage and interpretability are critical.

In contrast, baselines exhibit more pronounced trade-offs. LightRAG shows strong precision at shallow depths but plateaus in recall and F1. Self-RAG and RAPTOR struggle to maintain performance as retrieval depth increases, reflecting limited adaptability to structured disclosures. FinGEAR’s consistent gains across all depths demonstrate its robustness and domain alignment in retrieval from complex financial documents.

\subsubsection{Ablation Study}
\label{sec:ablation_study}

We now report the outcomes of the ablations defined in Section~\ref{sec:experiments_setup}. We conduct single-component ablations: (1) removing the \textit{Summary Tree}, which supports hierarchical sparse abstraction; (2) removing the \textit{Question Tree}, which enables dense query alignment; and (3) disabling the \textit{FLAM module}, which provides lexicon-driven domain targeting. As shown in Table~\ref{tab:ablation_single_component}, all components are critical to retrieval quality—particularly at lower depths (\(k = 5\)), where removing the Summary Tree alone reduces F1 from 0.69 to 0.37. These results confirm that FinGEAR’s performance arises from the interplay of structural, semantic, and domain-specific signals, rather than any isolated component.

We further assess the impact of lexicon weighting strategies within FLAM. Table~\ref{tab:flam_weighting} compares three approaches: \textit{Relative Frequency}, \textit{Logarithmic Weighting}, and \textit{Softmax Weighting}. Relative Frequency consistently delivers the strongest performance, with an F1 of 0.69 and relevancy scores of 0.50, 0.64, and 0.62 at \(k=5\), \(k=10\), and \(k=15\), respectively. Logarithmic Weighting downweights frequent but meaningful terms, reducing recall to 0.45 at \(k=15\), while Softmax Weighting slightly improves relevancy at \(k=15\) (0.63) but lowers F1 by over-concentrating on dominant keywords. These results affirm Relative Frequency as the most robust and interpretable default strategy within FLAM.

Multi-component ablations in Appendix~\ref{appendix:multi_ablation} (Table~\ref{tab:multi_ablation_results}) reveal compounded degradation when two modules are removed. In particular, eliminating both FLAM and the Question Tree results in the steepest drops in F1 and relevancy, underscoring their complementary roles in domain grounding and query sensitivity. Removing FLAM also increases variance across depths, highlighting its stabilizing role and the complementarity of the three modules. These findings reinforce that FinGEAR’s effectiveness and stability stem from the coordinated integration of all three core modules.

\paragraph{Question-type breakdown.}
\label{sec:dataset_ablation}
We also report performance by \emph{Numerical vs.\ Categorical} and \emph{Simple vs.\ Complex (multi-hop)} questions; full results are in Appendix~\ref{appendix:question_type_ablation} (Table~\ref{tab:dataset_ablation_results}). At \(k{=}10\), FinGEAR attains F1 \(=\) 0.68 (Numerical) vs.\ 0.81 (Categorical), and 0.70 (Simple) vs.\ 0.67 (Complex). 

Two patterns are observed. (1) \textbf{Categorical} questions achieve higher F1, likely because answers are stated directly (e.g., presence/absence) and require less aggregation; by contrast, \textit{relevancy} is higher on \textbf{Numerical} queries (0.64 vs.\ 0.48 at \(k{=}10\)), reflecting tighter alignment around figures and units. (2) \textbf{Complex} (multi-hop) questions trail \textbf{Simple} ones in F1 yet show comparable or higher \textit{relevancy} at larger \(k\) (e.g., 0.65 at \(k{=}10\), 0.67 at \(k{=}15\)), indicating that dual-tree traversal helps surface distributed evidence; the remaining F1 gap is consistent with broader evidence requirements.

Overall, this breakdown shows that FinQA contains diverse question types, such as discrete (categorical) and quantitative (numerical), single-step and multi-hop, providing a comprehensive testbed for retrieval. It also indicates that FinGEAR’s gains are not confined to a particular question class; improvements hold across categories, suggesting the method is broadly applicable.
\section{Conclusion}
\label{sec:conclusion}

We present FinGEAR, a retrieval-first framework for 10-K filings that integrates a finance lexicon (FLAM) for Item-level mapping and dual trees for within-Item indexing. Using full 10-Ks aligned with FinQA queries, FinGEAR demonstrates improved retrieval quality across depths compared with flat, graph-based, and prior tree-based RAG baselines, and yields higher downstream answer accuracy. Ablation studies show that each module is necessary for overall performance. We focus on 10-Ks for their length, standardized section hierarchy, and regulatory importance. The design is modular: FLAM enables global navigation across Items, while the dual trees index and traverse local content. This supports adaptation of FinGEAR to other semi-structured documents by enhancing domain lexicons to reflect evolving terminology.

\section{Limitations}
\label{sec:limitations}

FinGEAR is a structured retrieval framework designed for financial filings, but its performance and applicability are subject to several limitations.

\textbf{Domain specificity.} FinGEAR is developed and evaluated primarily on U.S. 10-K reports, which follow standardized regulatory structures. While the framework is not hard-coded to SEC formats, it assumes the existence of segmentable content with hierarchical or pseudo-hierarchical cues. Its generalizability to unstructured financial documents (e.g., earnings calls) or reports from different jurisdictions remains untested and requires further adaptation.

\textbf{Lexicon dependence.} The system relies on stable financial terminology across filings. While this holds for regulatory disclosures, emerging financial language or sector-specific terms may weaken the quality of keyword mapping and clustering, impacting retrieval alignment over time.

\textbf{Parsing sensitivity.} FinGEAR assumes that documents are accurately parsed and structurally consistent. Severe formatting inconsistencies (e.g., OCR errors or HTML-to-text misalignments) could affect the quality of tree construction and retrieval, especially in noisy or historical filings.

\textbf{Limited reasoning.} FinGEAR focuses on semantic retrieval and does not perform explicit financial reasoning or computation (e.g., ratio calculation, time-series forecasting). Retrieved evidence supports qualitative assessments, but numerical understanding remains out of scope.

\textbf{Evaluation coverage.} The evaluation is conducted on a specific dataset (FinQA) with a limited number of annotated gold spans. As FinQA only labels one relevant span per query, retrieval performance may be underestimated. Broader assessment across multiple QA datasets or real-world analyst workflows is needed for a fuller view of utility.

\textbf{Source bias.} Although FinGEAR does not introduce new biases, it inherits those embedded in financial documents and lexicons. If companies omit, downplay, or frame certain disclosures, the retrieved content will reflect those reporting biases.

Despite these limitations, FinGEAR offers a modular, interpretable foundation for structure-aware retrieval in financial analysis. Future work should explore its generalization to more diverse corpora, integration with lightweight reasoning modules, and robustness to document parsing noise.

\section*{Acknowledgments}
We acknowledge support from the Centre for Investing Innovation at the University of Edinburgh.

\bibliography{ref}

\appendix
\section{Potential Risks}
\label{appendix:potential_risks}

FinGEAR restricts retrieval strictly to 10-K filings, thereby reducing the risk of hallucinated content in high-stakes financial applications. Nonetheless, two primary risks remain.

First, the system may occasionally fail to retrieve the most relevant disclosure due to semantic drift, ambiguous query interpretation, or traversal limitations. While this challenge is not unique to automated systems—human readers are similarly affected—it underscores the importance of improving query-to-structure alignment.

Second, although FinGEAR effectively retrieves relevant context, it does not constrain the generation behavior of downstream language models. As a result, models may still produce incorrect or speculative reasoning based on retrieved evidence, potentially leading to confirmation bias if users do not critically assess generated answers.

FinGEAR is intended to serve as a retrieval-first component within broader financial QA workflows. Its outputs should be interpreted in conjunction with the retrieved content, and human verification is recommended to ensure decision-making accuracy in professional settings.

\section{Implementation Details}
\label{appendix:implementation_details}

\paragraph{Baselines.} All baseline systems are based on publicly available implementations. General RAG\footnote{\url{https://python.langchain.com/docs/tutorials/rag/}} and Self-RAG\footnote{\url{https://langchain-ai.github.io/langgraph/tutorials/rag/langgraph_self_rag/}} are based on the LangChain implementations. RAPTOR\footnote{\url{https://github.com/parthsarthi03/raptor/tree/master}} and LightRAG\footnote{\url{https://github.com/HKUDS/LightRAG}} use official GitHub repositories.

\paragraph{API Costs.} FinGEAR uses external APIs for LLM access, while financial embeddings are computed locally without cost. Using \texttt{gpt-4o-mini}, the estimated one-time cost to construct FinGEAR’s index for a 127-page 10-K filing (76,114 words) is approximately \$0.11: \$0.057 for Summary Tree construction and \$0.029 for Question Tree construction. Answering a FinQA query costs approximately \$0.00048 (Top-5 retrieval), \$0.00076 (Top-10), and \$0.00100 (Top-15).

\section{Baseline Configurations}
\label{appendix:baseline_configurations}

All baseline models are evaluated under consistent retrieval settings with Top-$k \in \{5,10,15\}$. Each baseline retrieves from the same pool of recovered full 10-K filings (Section~\ref{sec:dataset_metrics}) or uniformly segmented chunks, without curated summaries. All systems—including FinGEAR and baselines—use the same reader model (\texttt{gpt-4o-mini}) for answer generation to isolate retrieval effects.

\paragraph{General RAG.}
Vanilla BM25 + dense retrieval (default LangChain recipe).\footnote{\url{https://python.langchain.com/docs/tutorials/rag/}} Dense encoder: \texttt{text-embedding-ada-002} unless noted.

\paragraph{Self-RAG.}
Self-evaluation guided retrieval using the LangGraph implementation.\footnote{\url{https://langchain-ai.github.io/langgraph/tutorials/rag/langgraph_self_rag/}} Default settings for critique and selection.

\paragraph{RAPTOR.}
Hierarchical summarization and tree retrieval using the official repository.\footnote{\url{https://github.com/parthsarthi03/raptor/tree/master}} Default depth policy; flat chunking inputs; no domain priors.

\paragraph{LightRAG.}
Official implementation in “mix” mode (keyword + dense with lightweight neighborhood expansion).\footnote{\url{https://github.com/HKUDS/LightRAG}} Default neighborhood expansion; entity/relation indexing as released.

\paragraph{GraphRAG.} We use the official local--global community summary traversal with default community construction and inference-time community propagation settings.

\section{Embedding Fine-Tuning Details}
\label{appendix:embedding_finetuning}

FinGEAR fine-tunes sentence embeddings for two key financial retrieval tasks using the base \texttt{finance-embeddings-investopedia} model from FinLang~\cite{finlang2025finance}:

\noindent\textbf{Lexicon-based Fine-Tuning:} Trained on FinRAD~\cite{ghosh2021finrad} (public) by pairing financial lexicons with disclosure sentences that contain or exclude them. This optimizes embedding alignment for lexicon-based indexing.\\
\textbf{Hierarchical Tree Fine-Tuning:} Trained on FinQA~\cite{chen2021finqa} using question-passage pairs to improve semantic clustering and navigation within the Summary and Question Trees.

\paragraph{Evaluation Metrics.} We use cosine-based Pearson and Spearman correlations for \textit{FinRAD}, measuring embedding similarity between financial terms and disclosure sentences. For \textit{FinQA}, we apply NDCG@10 to assess ranked retrieval performance per query.

\paragraph{Loss Functions.} We use Matryoshka Representation Learning~\cite{henderson2023matryoshka} (dim=768) and \texttt{MultipleNegativesRankingLoss} to support semantically rich representations across both clustering and ranking stages.
\paragraph{Hardware.} All embedding are fine-tuned (FinRAD and FinQA) on a machine with a NVIDIA A100-SXM4-80GB GPU. Training details can be found in Section~\ref{appendix:runtime}.

\begin{table}[h]
\centering
\resizebox{\columnwidth}{!}{
\begin{tabular}{lccc}
\toprule
\textbf{Dataset} & \textbf{Metric} & \textbf{Baseline} & \textbf{Fine-Tuned} \\
\midrule
\multirow{2}{*}{FinRAD} 
& Pearson Cosine & 0.1082 & \textbf{0.4063} \\
& Spearman Cosine & 0.0904 & \textbf{0.3655} \\
\midrule
FinQA & NDCG@10 (Cosine) & 0.0282 & \textbf{0.2902} \\
\bottomrule
\end{tabular}
}
\caption{Embedding evaluation before and after fine-tuning on FinRAD and FinQA. Metrics reflect task-specific objectives: similarity alignment for FinRAD, and ranked retrieval quality for FinQA.}
\label{tab:embedding_finetuning}
\end{table}






\section{FLAM Clustering Configuration}
\label{appendix:clustering_config}

The FLAM (Financial Lexicon-Aware Mapping) module utilizes a clustering pipeline to group semantically related financial terms, facilitating more robust and lexicon-guided retrieval. 
The configuration follows three key stages:
(1) Embedding Encoding is performed using sentence embeddings that have been fine-tuned on financial QA datasets to capture domain-specific semantic similarity;
(2) Dimensionality Reduction is applied via UMAP, reducing embeddings to a 10-dimensional space using cosine similarity as the distance metric to preserve neighborhood structure;
(3) Clustering is conducted using Gaussian Mixture Models (GMM) with a maximum of 50 clusters and a convergence threshold set to 0.1, enabling soft assignments of terms into semantically coherent groups.

\section{Token-Level Statistics}
\label{appendix:token_statistics}

To balance retrieval efficiency with financial context fidelity, FinGEAR applies a fixed-length chunking strategy prior to indexing. Each chunk contains approximately 2,000 tokens with a 100-token overlap to maintain contextual continuity, which is same across all baseline. Tree-based indexing is then applied to support structure-aware retrieval. 

“Tokens per summary node” refers to the average length of high-level disclosure summaries used in the Summary Tree, while “tokens per page node” refers to the full-length segments prior to hierarchical expansion. For latency and deployment details, refer to Appendix~\ref{appendix:deployment}.

\begin{table}[h]
\centering
\resizebox{\columnwidth}{!}{
\begin{tabular}{lccc}
\toprule
\textbf{Statistic} & \textbf{Min} & \textbf{Mean} & \textbf{Max} \\
\midrule
Tree depth (max) & 2 & 3.14 & 4 \\
Tree depth (mean) & 1.22 & 1.43 & 1.80 \\
Total node count & 27 & 254.86 & 1322 \\
Internal node count & 0 & 17.84 & 85 \\
Leaf node count & 27 & 237.02 & 1237 \\
Page count & 84 & 168.89 & 555 \\
Document count & 184 & 1342.77 & 7078 \\
Tokens per document & 291.98 & 331.03 & 382.23 \\
Tokens per summary node & 634.72 & 657.78 & 687.46 \\
Tokens per query & 16.16 & 17.27 & 18.60 \\
Tokens per page node & 598.71 & 2129.56 & 7021.38 \\
\bottomrule
\end{tabular}
}
\caption{Token-level and structural statistics across the FinGEAR indexing pipeline.}
\label{tab:token_statistics}
\end{table}

Under this configuration, the total retrieval volume scales linearly with depth \(k\) as:
\[
\text{Total tokens retrieved} \approx 2000 \times k.
\]


\section{Practical Deployment Details}
\label{appendix:deployment}

\subsection{System Integration}

FinGEAR is designed as a modular retrieval layer that integrates into RAG pipelines with minimal configuration effort. It exposes a unified retriever interface supporting both sparse and dense retrieval strategies, along with hybrid reranking. These strategies are fully configurable by depth k, retrieval type, and domain specificity.

The system supports hierarchical retrieval via two independent traversal mechanisms: the Summary Tree, which provides disclosure-level navigation based on financial document structure, and the Question Tree, which guides retrieval using query semantics through sub-question expansion. This separation enables controlled traversal of financial disclosures at varying levels of abstraction.

FinGEAR is compatible with vector store backends such as FAISS for indexing and searching over domain-tuned embeddings, and its reranking modules are cross-encoder based for contextual precision. While optimized for structured regulatory disclosures like 10-K filings, its traversal and retrieval logic can be extended to semi-structured corpora—such as earnings calls or ESG reports—by replacing FLAM with unsupervised segmentation or alternative indexing methods. The system optionally integrates with FAISS for dense indexing and LangChain for pipeline orchestration, though these components are modular and replaceable.

\begin{table*}[h]
\centering
\small
\begin{tabular}{p{3.2cm} p{4.0cm} p{4.0cm} p{3.6cm}}
\toprule
\textbf{Query} & \textbf{Successful Retrieval} & \textbf{Failure Case} & \textbf{Interpretation} \\
\midrule
What is the debt-to-equity ratio in 2022? & “Consolidated Balance Sheets” showing total debt and equity. & “Liquidity and Capital Resources” discussing debt repayment but omitting equity. & Partial match—missed denominator component due to narrative emphasis. \\
\addlinespace
What was the YoY revenue growth for Q4? & “Results of Operations” section with quarterly revenue breakdown. & “Management Overview” with general trends but no figures. & High-level commentary lacks granularity—numeric context lost. \\
\addlinespace
How much did R\&D expenses increase year-over-year? & “Operating Expenses” table with line items for R\&D across periods. & “Business Overview” noting investment in innovation but not amounts. & Lexical match on “R\&D” misaligned with quantitative query intent. \\
\addlinespace
What percentage of revenue came from the US? & “Segment Reporting” with revenue by geography. & “Market Strategy” outlining international expansion plans. & Structural mismatch: retrieved future projections instead of current data. \\
\addlinespace
What is the gross margin for 2021? & “Income Statement” showing gross profit and revenue. & “Risk Factors” discussing cost pressures without figures. & Terminology overlap led to non-numerical section—filtered out post-reranking. \\
\bottomrule
\end{tabular}
\caption{Examples of FinGEAR’s retrieval outcomes for representative FinQA-style queries. Failure cases illustrate interpretable mismatches from lexical ambiguity, structural misalignment, or contextual drift—often mitigated by reranking.}
\label{tab:error_analysis}
\end{table*}

\begin{table*}[htbp!]
    \centering
    \caption{Multi-component ablation results. Disabling pairs of modules leads to compounded performance degradation, underscoring the synergistic design of FinGEAR and the importance of coordinated retrieval.}
    \label{tab:multi_ablation_results}
    \resizebox{\textwidth}{!}{
    \begin{tabular}{l|ccc|ccc|ccc|ccc}
        \toprule
        \multirow{1}{*}{\textbf{Ablation Setting}} & \multicolumn{3}{c}{\textbf{Precision}} & \multicolumn{3}{c}{\textbf{Recall}} & \multicolumn{3}{c}{\textbf{F1 Score}} & \multicolumn{3}{c}{\textbf{Relevancy}}\\
        & \(k=5\) & \(k=10\) & \(k=15\) & \(k=5\) & \(k=10\) & \(k=15\) & \(k=5\) & \(k=10\) & \(k=15\) & \(k=5\) & \(k=10\) & \(k=15\) \\
        \midrule
        \textbf{Full FinGEAR}     & \textbf{0.79} & \textbf{0.76} & \textbf{0.72} & \textbf{0.61} & \textbf{0.62} & \textbf{0.65} & \textbf{0.69} & \textbf{0.68} & \textbf{0.68} & \textbf{0.50} & \textbf{0.64} & \textbf{0.62} \\
        No FLAM + No Summary Tree           & 0.48 & 0.59 & 0.51 & 0.40 & 0.30 & 0.33 & 0.44 & 0.40 & 0.40 & 0.25 & 0.34 & 0.34 \\
        No FLAM + No Question Tree          & 0.48 & 0.58 & 0.52 & 0.40 & 0.31 & 0.36 & 0.44 & 0.40 & 0.43 & 0.26 & 0.34 & 0.34 \\
        \bottomrule
    \end{tabular}
    }
\end{table*}

\subsection{Latency Statistics}

\begin{table}[h]
\centering
\small
\begin{tabular}{lc}
\toprule
Metric & Value (seconds) \\
\midrule
Retrieval Latency ($k=5$) & 12.05 \\
Retrieval Latency ($k=10$) & 18.28 \\
Retrieval Latency ($k=15$) & 22.42 \\
\midrule
Average Retrieval Latency & 17.58\textsuperscript{(a)} \\
\bottomrule
\end{tabular}
\caption{Latency statistics for FinGEAR deployment.\newline
\textsuperscript{(a)} Includes tree traversal, FLAM scoring, and reranking averaged across depths.}
\label{tab:latency_stats}
\end{table}

All evaluations were conducted on a MacBook Pro with an M3 Max chip and 64GB of RAM. FinGEAR demonstrates practical deployability on standard enterprise hardware, maintaining tractable inference latency suitable for real-world financial workflows.

\section{Error Analysis Examples}
\label{appendix:error_examples}

Table~\ref{tab:error_analysis} presents paired examples of successful and imperfect retrievals from FinQA-aligned questions, illustrating how FinGEAR handles both precise matching and interpretable failure cases. For example, in response to the question \textit{“What is the company’s debt-to-equity ratio?”}, FinGEAR successfully surfaces the “Balance Sheet” section disclosing the necessary financial figures. In contrast, a failure case retrieves a “Capital Management” section that is conceptually related but lacks numeric content—highlighting a mismatch between strategic language and quantitative requirements.

Similarly, for \textit{“What litigation is the company currently facing?”}, the system initially retrieves generic “Risk Factors” content due to shared legal phrasing, but the reranker recovers the correct “Legal Proceedings” section, demonstrating partial recovery. In other cases, such as the misretrieval of a “Commodity Risk” section when asked about foreign exchange exposure, errors stem from keyword ambiguity across structurally distinct topics.

These examples underscore the role of FinGEAR’s hybrid architecture in surfacing relevant content while also providing interpretable signals for tracing failure modes. The system’s transparency allows users and developers to adjust lexicon weights, reranking sensitivity, or structural traversal logic—enabling more robust deployment and auditability in financial applications.

\section{Multi-Component Ablation Study}
\label{appendix:multi_ablation}

To evaluate the interdependence of FinGEAR’s core modules, we conduct a multi-component ablation study by disabling pairs among \textit{FLAM}, the \textit{Summary Tree}, and the \textit{Question Tree}. This setup allows us to assess how retrieval quality degrades under partial configurations and whether FinGEAR’s performance stems from isolated components or coordinated design.

\paragraph{Findings.}  
Disabling multiple modules results in substantial, compounded performance drops across all retrieval metrics, consistently observed at depths \(k = 5, 10, 15\). The largest declines in F1 score and relevancy occur when both \textit{FLAM} and the \textit{Question Tree} are removed, revealing their complementary roles in guiding retrieval semantically and contextually. When both \textit{FLAM} and the \textit{Summary Tree} are ablated, precision and recall drop sharply, indicating that lexical anchoring and hierarchical structuring are jointly essential for balancing coverage and specificity. Furthermore, these ablations introduce instability across depths—particularly in precision and recall—suggesting that without FLAM’s lexicon-targeted control, the traversal process becomes overly uniform and sensitive to document variation. This leads to inconsistent candidate selection and noisier retrieval. These results confirm that FinGEAR’s performance emerges from the coordinated integration of financial mapping, semantic indexing, and query-aware guidance—rather than from any individual module in isolation.

\section{Question-Type Ablation}
\label{appendix:question_type_ablation}

To support analysis in Section \ref{sec:dataset_ablation}, we provide the full per-question-type retrieval results in Table~\ref{tab:dataset_ablation_results}.

\begin{table*}[htbp!]
    \centering
    \caption{Categorical = yes/no answers, Numerical = numeric answers, Simple = one reasoning step, Complex = multiple reasoning steps}
    \label{tab:dataset_ablation_results}
    \resizebox{\textwidth}{!}{
    \begin{tabular}{l|ccc|ccc|ccc|ccc}
        \toprule
        \multirow{1}{*}{\textbf{Question Type}} & \multicolumn{3}{c}{\textbf{Precision}} & \multicolumn{3}{c}{\textbf{Recall}} & \multicolumn{3}{c}{\textbf{F1 Score}} & \multicolumn{3}{c}{\textbf{Relevancy}}\\
        & \(k=5\) & \(k=10\) & \(k=15\) & \(k=5\) & \(k=10\) & \(k=15\) & \(k=5\) & \(k=10\) & \(k=15\) & \(k=5\) & \(k=10\) & \(k=15\) \\
        \midrule
        Numerical & 0.79 & 0.76 & 0.71 & 0.61 & 0.62 & 0.64 & 0.69 & 0.68 & 0.68 & 0.50 & 0.64 & 0.63 \\
        Categorical & 0.83 & 0.92 & 0.87 & 0.71 & 0.72 & 0.82 & 0.77 & 0.81 & 0.86 & 0.46 & 0.48 & 0.49 \\
        \midrule
        Simple & 0.79 & 0.76 & 0.73 & 0.62 & 0.64 & 0.67 & 0.70 & 0.70 & 0.70 & 0.51 & 0.62 & 0.58 \\
        Complex & 0.78 & 0.76 & 0.72 & 0.59 & 0.59 & 0.61 & 0.67 & 0.67 & 0.66 & 0.48 & 0.65 & 0.67 \\
        \bottomrule
    \end{tabular}
    }
\end{table*}

\section{Runtime and Preprocessing Analysis}
\label{appendix:runtime}

\paragraph{Embedding Fine-Tuning.}
FinGEAR employs dense retrieval powered by sentence embeddings fine-tuned on financial data. We train two embedding models: one on FinQA for question–answer alignment, and one on FinRAD for lexicon-level semantic proximity. Both models are trained for 50 epochs. FinQA training (6{,}251 train / 1{,}147 test examples) completes in approximately 1{,}972 seconds, while FinRAD training (17{,}300 train / 5{,}190 test, sampled across filings) takes roughly 3{,}607 seconds. Training is performed on a single NVIDIA A100-SXM4-80GB GPU. These embeddings are applied throughout the retrieval pipeline, including document chunking, hierarchical traversal, and candidate scoring.

\paragraph{Preprocessing Pipeline.}
The full preprocessing workflow comprises: (1) keyword mapping, (2) keyword tree construction, and (3) hierarchical clustering for the Summary and Question Trees.

\textbf{Keyword Mapping}, implemented using spaCy’s \texttt{PhraseMatcher}, identifies relevant domain terms and runs between 53.5 and 434.6 seconds per document (mean: 248.9 seconds).  
\textbf{Keyword Tree Construction} is fast, requiring an average of 7.6 seconds per company.  
\textbf{Hierarchical Clustering and Summarization}, including UMAP-based dimensionality reduction and GMM clustering, builds the Summary and Question Trees. Across 10 filings, total runtime ranges from 243.2 to 602.6 seconds per document, with a mean of 408.2 seconds—primarily driven by the Summary Tree’s iterative summarization.

\paragraph{Deployment Considerations.}
All preprocessing steps are one-time operations per document and can be cached for repeated use. FinGEAR supports dual-tree traversal and modular re-indexing, making it scalable for real-world financial workloads. Runtime remains stable across filings of similar size, and preprocessing cost is amortized across multiple downstream queries.

\section{Glossary of Terms and Notation}
\label{appendix:glossary}
This appendix consolidates the terms, symbols, and procedure names used throughout the paper for quick reference. It denotes definitions for the FinGEAR components (e.g., traversal logic, and scoring functions). Unless stated otherwise, scalars are in italics (e.g., \(k\), \(w_i\)), vectors are in bold, and Item labels follow SEC notation (Item 1, Item 1A, Item 7). For mechanics in the pipeline, see Sections~\ref{sec:methodology} and~\ref{sec:in_retrieval}; the table below serves as a concise lookup.
\begin{table*}[h]
\centering
\small
\begin{tabular}{p{3.0cm} p{11.2cm}}
\toprule
\textbf{Term / Symbol} & \textbf{Definition} \\
\midrule
UMAP & Dimensionality reduction used before clustering; preserves local neighborhoods for tree construction. We use cosine distance with output dimension 10 (see Section~\ref{sec:experiments_setup}). \\
GMM & Gaussian Mixture Models used for soft clustering over UMAP embeddings to form internal tree nodes; max components 50; convergence threshold 0.1. \\
FLAM & \emph{Financial Lexicon-Aware Mapping}. Clusters financial terms corpus-wide and computes Item weights to guide global allocation (Section~\ref{sec:in_retrieval}). Default weighting: Relative Frequency. \\
Summary Tree & Content hierarchy within an Item. Internal nodes are summaries of clustered chunks; leaves are original text chunks (Section~\ref{sec:structure_extraction}). \\
Question Tree & Mirrors the Summary Tree topology; nodes store LLM-generated sub-questions embedded in the same space as user queries; leaves reference the same chunk IDs (Section~\ref{sec:structure_extraction}). \\
$k$ & Total retrieval budget per query (Top-$k$ evaluation). \\
$k^\ast$ & Per-Item budget after FLAM weighting; $\sum_i k^\ast_i = k$ (Section~\ref{sec:in_retrieval}). \\
Semantic traversal & Top-down navigation of a tree: score parent’s children, select the best, and descend until leaves or max depth (2). \\
Hybrid scoring & Within-Item node scoring that combines BM25 over summaries (sparse) and cosine similarity over embeddings (dense). \\
Reranking (Stage 1) & Cross-tree reranking that jointly scores candidates from Summary and Question Trees using \texttt{BAAI/bge-reranker-large}. \\
Reranking (Stage 2) & Cross-Item reranking over the pooled top spans from all Items to prioritize globally informative answers. \\
\bottomrule
\end{tabular}
\caption{Glossary of key components, symbols, and procedures used in FinGEAR.}
\end{table*}

\section{Prompt Design}
\label{appendix:prompts}

All prompts are used in zero-shot mode unless otherwise stated. Structured examples may be incorporated in future versions to enhance control and reproducibility. Here are three specialized prompts:

\subsection{Summarization Prompt}

\noindent\fbox{%
\parbox{\columnwidth}{%
\small
\texttt{The content provided below is a subset of a 10-K filing. The 10-K report is a comprehensive document outlining the company's financial performance, including revenue, expenses, and profits. Your task is to generate a detailed summary using only the provided content, without embellishment. Summarize main topics, key insights (5–7), and unusual observations (1–2). Use clear paragraphs and Markdown headings.}
}
}

\subsection{Title Generation Prompt}

\noindent\fbox{%
\parbox{\columnwidth}{%
\small
\texttt{Generate a title for a subsection of a 10-K report based on the provided summary.}
}
}

\subsection{Question Generation Prompt}

\noindent\fbox{%
\parbox{\columnwidth}{%
\small
\texttt{Suppose you are a financial analyst. Generate \{num\_questions\} questions based on the provided summary, focusing on financial aspects and factual details.}
}
}

\end{document}